# Optimizing tiny colorless feedback delay networks

Gloria Dal Santo[1*], Karolina Prawda[1,2], Sebastian J. Schlecht[1,3] and Vesa Välimäki[1]

**Abstract**

A common bane of artificial reverberation algorithms is spectral coloration in the synthesized sound, typically manifesting as metallic ringing, leading to a degradation in the perceived sound quality. In delay network methods, coloration is more pronounced when fewer delay lines are used. This paper presents an optimization framework in which a tiny differentiable feedback delay network, with as few as four delay lines, is used to learn a set of parameters to iteratively reduce coloration. The parameters under optimization include the feedback matrix, as well as the input and output gains. The optimization objective is twofold: to maximize spectral flatness through a spectral loss while maintaining temporal density by penalizing sparseness in the parameter values. A favorable narrow distribution of modal excitation is achieved while maintaining the desired impulse response density. In a subjective assessment, the new method proves effective in reducing perceptual coloration of late reverberation. Compared to the author's previous work, which serves as the baseline and utilizes a sparsity loss in the time domain, the proposed method achieves computational savings while maintaining performance. The effectiveness of this work is demonstrated through two application scenarios where smooth-sounding synthetic room impulse responses are obtained via the introduction of attenuation filters and an optimizable scattering feedback matrix.

**Keywords** Audio systems, Gradient methods, Optimization, Psychoacoustics, Reverberation

## 1 Introduction

Delay-based recursive structures are common for generating artificial reverberation [1]. The initial approach incorporating delay lines and feedback loops was introduced by Schroeder and Logan in the early 1960s [2]. Over time, this structure underwent various modifications, eventually evolving into the feedback delay network (FDN) [3], which has since the 1990s become a widely adopted algorithm for digital reverberation synthesis [1, 4–6].

*Correspondence:
Gloria Dal Santo
gloria.dalsanto@aalto.fi
[1] Acoustics Lab, Department of Information and Communications Engineering, Aalto University, Espoo FI-02150, Finland
[2] School of Physics, Engineering and Technology, University of York, York YO10 5DD, UK
[3] Multimedia Communications and Signal Processing, University of Erlangen-Nuremberg, Erlangen DE-91058, Germany

An issue commonly encountered in delay-based artificial reverberation is coloration, often manifesting as audible metallic ringing in the resulting sound [2]. Coloration is detrimental to perceived sound quality, deviating from the ideal smooth reverb by introducing fluctuations and prominent resonances in the spectrum, particularly in the tail. In the early days of digital artificial reverberation, Schroeder and Logan [2] sought to achieve colorlessness by concatenating delay-line-based allpass filters. Jot and Chaigne [3] introduced a two-stage design process involving the creation of a lossless system followed by the introduction of delay-proportional attenuation. However, these efforts proved insufficient, necessitating additional considerations regarding the selection of parameters to effectively reduce or eliminate the coloration [2, 7–9].

Recently, the perceived coloration was attributed to the modal properties of synthesized reverbs, with a wide distribution of modal excitations linked to increased coloration [10]. Therefore, modal decomposition of artificial reverberation architectures, such as FDNs [11], can aid in identifying phenomena that are related to perceptual artifacts. In FDNs, the values of the modal excitations depend on all system parameters [11].

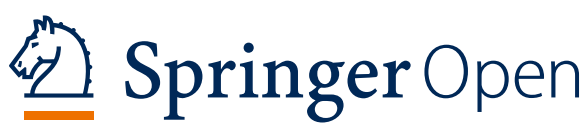

Another key factor in achieving natural-sounding reverb is the number and distribution of reflections over time, commonly referred to as echo density [9]. With fewer delay lines, typically fewer than 16, achieving a rapid build-up of reflections becomes more challenging. The feedback matrix, which governs energy distribution among delay lines, plays a crucial role in this process. When the feedback matrix is small, the energy distribution is less smooth due to less flexibility for adjusting spectral and temporal properties, resulting in noticeable periodicities or resonances that can sound unnatural.

In this paper, the objective is to optimize the gain parameters and feedback matrix of tiny FDNs, incorporating as few as four delay lines. Similarly to our previous work [12], the goal is to minimize the perceptual coloration observed in the impulse response (IR) by adjusting the parameters using stochastic gradient descent. In [12], we targeted the flatness of the magnitude response and temporal density by incorporating two loss functions in the frequency and time domains, respectively. For the time domain, the inverse discrete Fourier transform (DFT) operation was needed, which presented a bottleneck that we address in this paper.

The current work also demonstrates practical applications of the proposed approach in synthesizing natural-sounding late reverberation by optimizing the scattering feedback matrix and including in the FDN structure attenuation filters that benefit from the optimization. As we now concentrate solely on the frequency domain, the computational requirements are streamlined by analyzing the frequency response sparsely using batch processing over subsets of frequency samples. We attain control over temporal density by utilizing its correlation with the density of the feedback matrix entries instead of computing the system's IR. Evaluation is conducted by analyzing the modal excitation distribution. A perceptual evaluation against the FDN design before optimization shows that the proposed method successfully decreased perceived coloration.

This work distinguishes itself from recent studies presented in [13, 14], and their extension to the multiple-input and multiple-output case [15], by addressing issues underlying reverb synthesis using FDNs with few delay lines and by validating our approach on both objective and subjective tests. While Mezza et al. propose a fully optimizable FDN aimed to fit a target IR [13], our approach focuses exclusively on optimizing the gain and feedback matrix of an FDN, targeting the coloration of its IR. Delay lines can have a strong impact on the coloration and echo density. To mitigate their impact, we follow the studies in [4, 9, 16] to produce non-degenerating delay patterns. Furthermore, while [13–15] rely solely on objective validation of the results, our method is supported by formal listening tests to demonstrate its perceptual fidelity.

The proposed optimization aims to reduce the remaining audible artifacts present in an FDN, given a fixed set of delay lines, making the resulting sound more natural. At the same time, our method can be extended to IR synthesis tasks by incorporating a filter design of choice. Nonetheless, the optimization does not depend on the filter used, unlike [14], thereby opening a wider range of applications. Another key distinction is that our optimization method operates in the frequency domain, unlike the time-domain approach by Mezza et al. [13, 14], which entails higher computational complexity and longer training times.

The paper is structured as follows. In Sect. 2, we provide background information on FDNs, including their modal decomposition, and discuss coloration in the IR of FDNs. Section 3 introduces our proposed method for designing colorless FDNs, highlighting the improvements made compared to our previous work. Practical applications of this approach are demonstrated in Sect. 4. The results of both objective and perceptual evaluations are presented in Sects. 5 and 6, respectively. Finally, Sect. 7 offers concluding remarks.

## 2 Background

In the following, we provide background information on the FDN and its fundamental properties. This includes an explanation of its modal decomposition and a discussion on coloration in FDNs, which are central concepts in the proposed method.

### 2.1 Feedback delay network

An FDN is a recursive system consisting of delay lines, a set of gains, and a feedback matrix through which the delay outputs are coupled to the delay inputs. Figure 1 presents an example of a single-input, single-output (SISO) FDN architecture. The transfer function of the FDN can be written as

$$H(z) = \frac{Y(z)}{X(z)} = \boldsymbol{c}^\top \left[ \boldsymbol{D_m}(z)^{-1} - \boldsymbol{A} \right]^{-1} \boldsymbol{b} + d \,, \tag{1}$$

where $\boldsymbol{A}$ is the $N \times N$ feedback matrix when $N$ is the number of delay lines, $\boldsymbol{D_m}(z)$ is the $N \times N$ delay matrix, vectors $\boldsymbol{b}$ and $\boldsymbol{c}$ are $N \times 1$ column vectors of input and output gains, respectively, the scalar coefficient $d$ is the direct gain, and the operator $(\cdot)^\top$ denotes the transpose. The frequency-domain input and output signals are represented by $X(z)$ and $Y(z)$, respectively. The vector $\boldsymbol{m} = [m_1, \ldots, m_N]$ defines the lengths of delays in samples. Its corresponding delay matrix $\boldsymbol{D}_{\boldsymbol{m}}(z)$ is created by taking a diagonal matrix with entries given by $[z^{-m_1}, \ldots, z^{-m_N}]$. The sum of the delays gives the order of the system, i.e., $\mathcal{M} = \sum_{i=1}^{N} m_i$ [7].

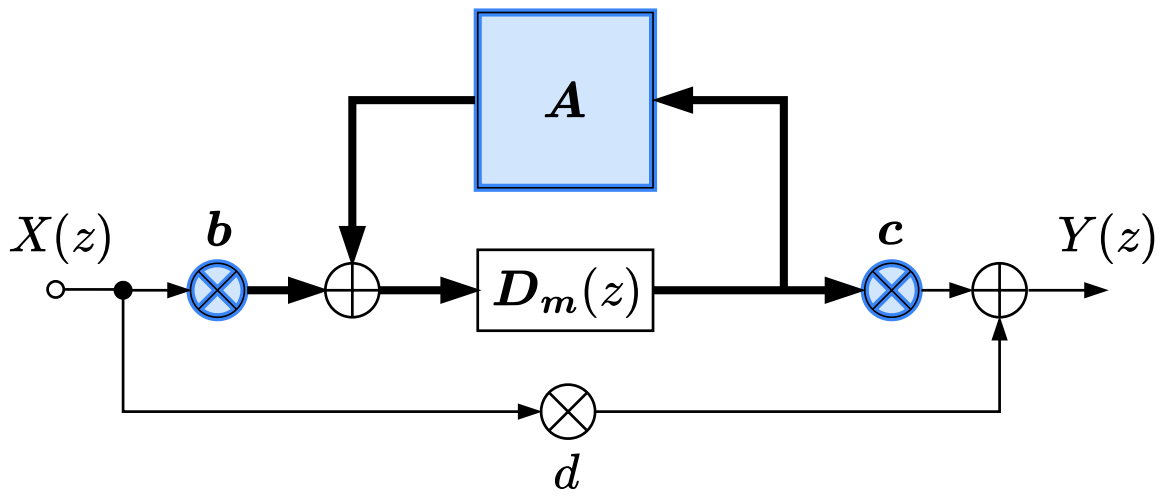

**Fig. 1** FDN with a single input, a single output, and an *N*-by-*N* feedback matrix ***A***. The thick lines indicate *N*-channel signal paths. The circled multiplication symbols indicate a vector multiplication. The parameters under optimization are highlighted in blue

The poles $\lambda_i$ of the system in (1) are the roots of its generalized characteristic polynomial (GCP) $p(z)$, and are fully characterized by $\boldsymbol{m}$ and $\boldsymbol{A}$:

$$p(z) = \det(\boldsymbol{D}_{\boldsymbol{m}}(z)^{-1} - \boldsymbol{A})\,. \tag{2}$$

With (2), the transfer function (1) can be expressed as a rational polynomial

$$H(z) = d + \frac{\boldsymbol{c}^\top \operatorname{adj}(\boldsymbol{D}_{\boldsymbol{m}}(z)^{-1} - \boldsymbol{A})\boldsymbol{b}}{p(z)}\,, \tag{3}$$

where $\operatorname{adj}(\cdot)$ indicates the matrix adjugate operator. This expression of the FDN transfer function can be used to compute the modal decomposition of the system directly. Moreover, the GCP provides explicit information regarding the stability of the system, which will subsequently be utilized for parameter-tuning purposes.

## 2.2 Modal decomposition

The IR of the FDN can be expressed as the sum of complex one-pole modes, also known as resonators, each defined by its pole $\lambda_i$ and its residue $\rho_i$ [11]:

$$h(n) = \sum_{i=1}^{\mathcal{M}} |\rho_i||\lambda_i|^n e^{\jmath(n\angle\lambda_i + \angle\rho_i)}\,,$$

where $|\cdot|$ denotes the magnitude, $\angle$ represents the argument of a complex number in radians, $\jmath = \sqrt{-1}$, and $n$ indicates the discrete time index.

The transfer function of the FDN (1) can be represented by its poles and residues through the partial fraction decomposition as

$$H(z) = d + \sum_{i=1}^{\mathcal{M}} \frac{\rho_i}{1 - \lambda_i z^{-1}}\,, \tag{4}$$

commonly referred to as the modal decomposition of the FDN [11]. The excitation and initial phase of the $i^{\text{th}}$ mode are determined by the magnitude $|\rho_i|$ and phase $\angle\rho_i$, respectively, of its corresponding residue. The magnitude and phase of the $i^{\text{th}}$ pole, $|\lambda_i|$ and $\angle\lambda_i$, respectively, dictate its decay rate and frequency. A common approach to calculate the modal decomposition (4) from the rational polynomial form of the transfer function (3) involves determining the system poles $\lambda_i$ from the eigenvalues of the linearized feedback matrix [11].

An important aspect for the remainder of this paper is the distribution of the modal excitation values $|\rho_i|$. While the decay of the modes $|\lambda_i|$ is typically governed entirely by the target reverberation time $T_{60}$, the modal excitation remains largely unconstrained by design. The connection between the distribution of modal excitation with perceived coloration [10] is a key finding underlying the proposed optimization method.

## 2.3 Homogeneous decay in FDN

Designing an artificial reverberator with FDN often starts by creating a lossless prototype, having an energy-preserving feedback loop [17, 18]. The FDN is said to be lossless if the roots of the GCP (2) have magnitude equal to one, i.e., $|\lambda_i| = 1$ for all $i$s [19]. The advantage of initially designing a lossless FDN lies in the straightforward implementation of frequency-dependent decay that equally influences all system poles. This is achieved by extending every delay with a frequency-dependent attenuation filter to meet the specified reverberation time [3].

In this study, we mainly focus on the specific case of frequency-independent homogeneous decay. This refers to the case where all modes experience the same decay rate, i.e., $|\lambda_i| = \gamma$ for all $i$ s. In Sect. 4.1, we demonstrate how an FDN optimized as such can be employed as a lossless prototype for synthesizing reverb with frequency-dependent decay.

Homogeneous decay is achieved with a feedback matrix $\boldsymbol{A}$ being the product of a unilossless matrix $\boldsymbol{U}$ and a diagonal matrix $\boldsymbol{\Gamma}$, whose entries are delay-proportional absorption coefficients, $\boldsymbol{\Gamma} = \operatorname{diag}(\gamma^{\boldsymbol{m}})$, where the vector $\gamma^{\boldsymbol{m}}$ represents the $\gamma$ value raised to the power of each corresponding delay in $\boldsymbol{m}$. The feedback matrix can be expressed as

$$\boldsymbol{A} = \boldsymbol{U}\boldsymbol{\Gamma}\,. \tag{5}$$

A matrix $\boldsymbol{U}$ is unilossless if, regardless of the choice of delays $\boldsymbol{m}$, its eigenvalues are unimodular and its eigenvectors are linearly independent. A matrix $\boldsymbol{U}$ satisfying the unitary condition, $\boldsymbol{U}\boldsymbol{U}^H = \boldsymbol{I}$, is also unilossless [9, 20]. As $\boldsymbol{U}$ is unilossless, the modal decay is controlled entirely by the gain-per-sample parameter $\gamma$, where $0 \leq \gamma \leq 1$. The gain-per-sample in dB is

$$\gamma_{\text{dB}} = \frac{-60}{f_{\text{s}} T_{60}}, \tag{6}$$

where $f_{\text{s}}$ is the sampling rate in Hz.

### 2.4 Coloration in FDN

Natural late reverberation can be modeled as white noise with exponential decay, providing perceptually ideal smooth diffuse reverberation [21]. In artificial reverberation, achieving a white noise generator involves ensuring the allpass property is maintained. Various perceptual artifacts of recursive reverberators can be attributed, though not exclusively, to deviations from this ideal, which then result in unwanted coloration.

Schroeder and Logan [2] made the initial attempt to produce colorless artificial reverberation by establishing specific requirements for the reverberators in addition to a flat frequency response. Overlapping normal modes across all frequencies, equal $T_{60}$ values for each mode, sufficient echo density, lack of periodicity in the time domain, and no periodic or comb-like frequency responses were deemed necessary to achieve colorlessness [2]. Despite fulfilling the aforementioned conditions, however, the Schroeder series allpass did not attain complete colorlessness.

A recent study was conducted to understand further the role of modal excitation in late reverberation coloration [10]. Listening test results suggest that a narrow distribution of the modal excitation values $|\rho_i|$ tends to result in a perceptually white spectrum [10]. However, when a subset of modes exhibits high $|\rho_i|$ values relative to the mean of the system's modal excitation distribution, coloration starts to become noticeable [10]. We also observed this outcome in our earlier study, where we conducted a listening test focusing on the coloration of an FDN-based reverb [12]. In this test, we compared the reverb before and after applying an optimization method, which resulted in a narrower modal excitation distribution.

The literature indicates that more than 6000 modes are needed for an IR to be perceived as rather colorless [16]. Additionally, it was shown that for large values of $\mathcal{M}$, the modes of the FDNs are uniformly distributed [11], preventing additional coloration that usually results from clusters of modes. Nonetheless, a flat magnitude response and a uniform modal frequency distribution are insufficient to achieve colorlessness.

To comprehend the complexities associated with coloration in FDN, we can examine the comb-filter structure as a special case. Specifically, when the feedback matrix $\boldsymbol{A}$ is diagonal, the FDN adopts the configuration of a parallel comb-filter structure. In the case of a homogeneous FDN with a diagonal feedback matrix, the transfer function in (1) is analogous to a superposition of comb filters with the delay line in the feedforward path and the gain $\gamma_i^m$ in the feedback. Each filter has a transfer function

$$H_{\text{comb}_i}(z) = \frac{z^{-m_i}}{1 - \gamma^{m_i} z^{-m_i}}. \tag{7}$$

The contribution of each filter to the total energy of the response can be calculated as

$$\|H_{\text{comb}_i}(z)\|_2^2 = \int_0^{2\pi} \left|H_{\text{comb}_i}(e^{\jmath\omega})\right|^2 \text{d}\omega \tag{8}$$

$$= \frac{1}{1 - \gamma^{2m_i}}, \tag{9}$$

where $\omega$ is the angular frequency, and $\|\cdot\|_2$ denotes the $\ell^2$ norm.

For any given positive value of $\gamma$ smaller than 1, (9) represents an exponential function. For positive $m_i$, the function approaches infinity as $m_i$ values decrease and tends to one as $m_i$ values increase. For example, let us consider the case when the gain-per-sample is $\gamma = 0.9999$ ($T_{60} = 1.44$ s). In this scenario, the energy of the response from the comb filter with $m_1 = 200$ is 8.5 times greater than that with $m_2 = 10m_1 = 2000$. Specifically, $\|H_1(z)\|_2^2 = 25.5$, while $\|H_2(z)\|_2^2 = 3$. Fundamentally, shorter delays contribute more energy and produce strongly audible metallic-sounding comb peaks. In contrast, longer delays contribute less energy and tend to be masked by the more dominant comb filters [4]. The same conclusion can be derived through pole-zero analysis of (7). By definition, FDNs are networks of comb filters whose feedback paths are interconnected through a feedback matrix. Therefore, the reasoning above applies to FDNs with dense feedback matrices as well.

Figure 2 shows a section of the magnitude response of two FDNs with identical parameters but different delay-line lengths. Since the resonant frequencies are uniformly distributed across the frequency, [11] the choice of the observed frequency range is arbitrary and serves only for visualization purposes. The feedback matrix is diagonal with values alternating between +1 and −1. The input and output gains are unit vectors. The lengths of the long delay lines are [533, 723, 753, 943, 1130, 1154, 1418, 165 4] samples, whereas the corresponding short delay-line lengths are [53, 72, 75, 94, 113, 115, 141, 165] samples. The FDN with shorter delay lengths, approximately one-tenth of the longer delays, exhibits larger magnitude values and more distinct, louder resonances than the one

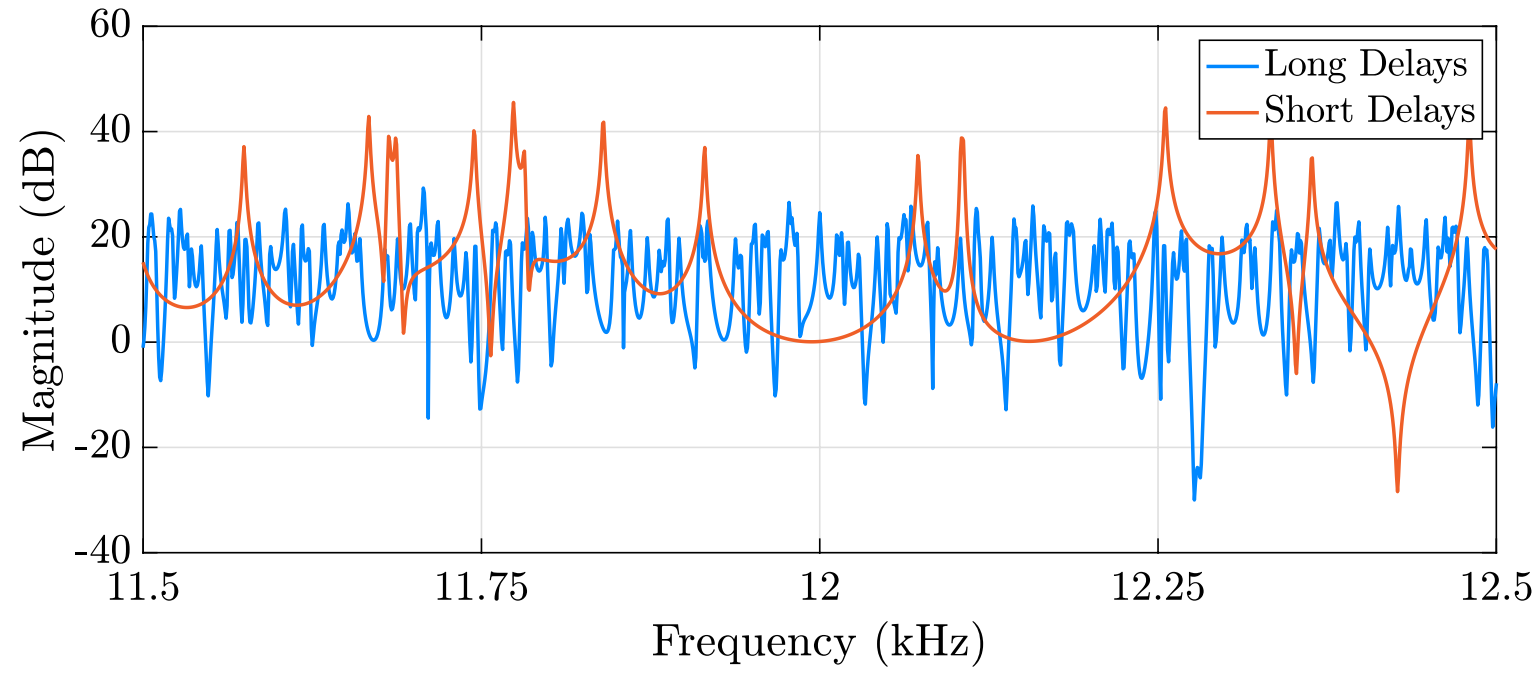

**Fig. 2** Magnitude response of a homogeneous FDN with size $N = 8$, diagonal matrix, and unit input and output gains, showcasing the difference between long and short delay lines when the former is 10 times longer than the latter

with long delay, which can be seen in Fig. 2. For long delays, the resonances become less noticeable, and the loudest frequencies exhibit a magnitude approximately 20 dB lower compared to the scenario with shorter delays. This example suggests that to achieve a colorless FDN, it is necessary to avoid strongly recirculating short delays and instead use long delay lines.

### 2.5 Problem statement

In this paper, our goal is to optimize the parameters of an FDN with a small number of delay lines, to enhance the perceptual colorlessness of the resulting IR. Given a set of delay lines, we optimize the feedback delay matrix $\boldsymbol{A}$, as well as its input and output gains $\boldsymbol{b}$ and $\boldsymbol{c}$. In Fig. 1, the FDN parameters that are optimized are indicated in blue.

Previous studies suggest that the choice of frequency-dependent attenuation has little impact on coloration [10]. Therefore, we conduct optimization on a long-ringing, frequency-independent prototype FDN. For ideal late reverberator synthesis with this setup, the spectrum's magnitude must remain constant across frequency and time. However, the recursive structure and the limited mixing capabilities of a tiny FDN introduce undesired coloration, which constitutes a primary challenge in our work, as emphasized in Sect. 2.4.

Another challenge is maintaining the temporal density of the IR during optimization. Our previous work demonstrated that focusing solely on the frequency-sampled magnitude response may lead to convergence towards a comb filter [12]. To address this efficiently, we utilize the relationship between temporal density and the density of the feedback matrix. For evaluation, we adopt modal decomposition of the FDN transfer function, as it has proven useful in analyzing the coloration of recursive systems [10]. To demonstrate the effectiveness of our method, we present various application scenarios that showcase its utilization for frequency-dependent attenuation and increased echo density, all while ensuring that the computational efficiency is maintained.

## 3 FDN optimization

In this section, we present a method to reduce coloration in the IR of an FDN with frequency-independent $T_{60}$. We employ stochastic gradient descent to optimize the gain parameters of the FDN, which has been made differentiable by utilizing the frequency sampling method.

### 3.1 Differentiable FDN

Using the frequency sampling method, the FDN is approximated as a finite impulse response (FIR) filter. An FIR filter is straightforward to implement, as it does not require a recursive structure. Moreover, an FIR filter can be easily parallelized and optimized for hardware acceleration, such as a graphics processing unit. To obtain the FIR approximation of $H(z)$, the delay matrix $\boldsymbol{D_m}(z)$ is evaluated at discrete frequency points

$$\boldsymbol{z}_M = \left[e^{\jmath\pi\frac{0}{M}}, e^{\jmath\pi\frac{1}{M}}, \ldots, e^{\jmath\pi\frac{M-1}{M}}\right], \tag{10}$$

where $M$ indicates the total number of frequency bins evenly distributed on the unit circle.

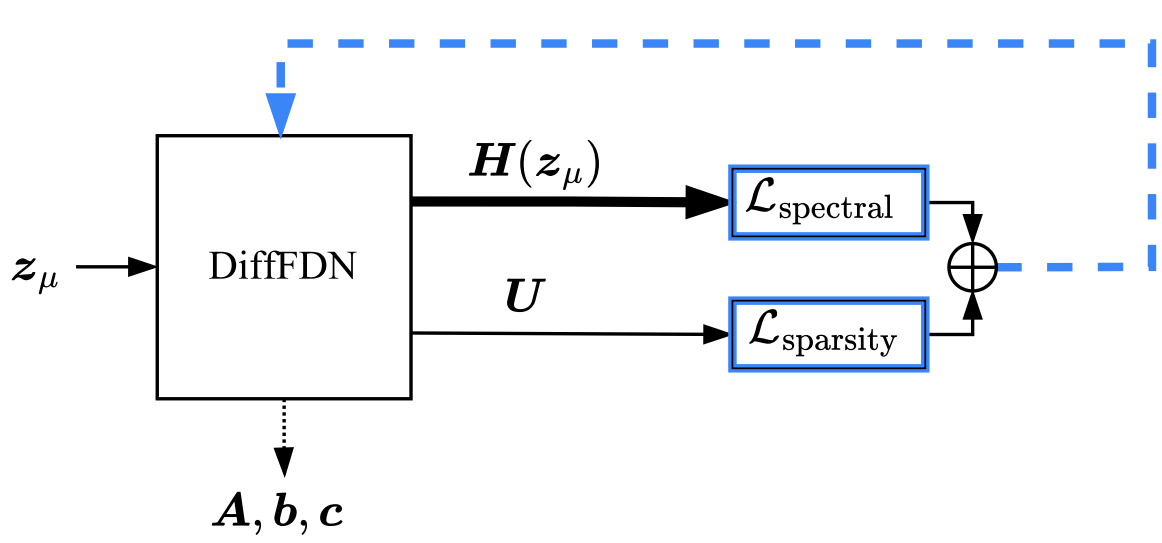

**Fig. 3** Architecture of the proposed optimization workflow. A thick line indicates an $N$-channel signal path. Backpropagation is highlighted with a dashed blue line

The block diagram of the architecture including a Differentiable FDN (DiffFDN) is shown in Fig. 3. We integrated $H(\boldsymbol{z}_M)$ into an optimization framework to estimate a set of FDN configurations based on a spectral loss and a sparsity loss by gradient descent. The learnable parameters are the feedback matrix $\boldsymbol{A}$ and the input and output gain vectors $\boldsymbol{b}$ and $\boldsymbol{c}$, respectively. The delay lengths $\boldsymbol{m}$ are set at initialization and kept constant during training. The direct gain $d$ is set to zero. During training, the FDN is set to have a homogeneous decay by forcing $\boldsymbol{A}$ to satisfy (5) for a given $\gamma$.

At each training step the estimated channel-wise transfer function $\boldsymbol{H}(z)$ is computed at a batch of frequency bins,

$$\boldsymbol{H}(\boldsymbol{z}_\mu) = \boldsymbol{c}\left[\boldsymbol{D}_{\boldsymbol{m}}(\boldsymbol{z}_\mu)^{-1} - \boldsymbol{A}\right]^{-1}\boldsymbol{b} + d\,. \tag{11}$$

Every batch is composed of $\mu < M$ different frequency points that are randomly sampled from $\boldsymbol{z}_M$. The value of $M$ is such that to ensure oversampling, we found empirically that $\mu$ can be as small as $0.05f_s$. At each epoch, $\boldsymbol{H}(z)$ is estimated at all frequency points within the vector $\boldsymbol{z}_M$. In our prior study [12], the computation of $\boldsymbol{H}(z)$ during each training step involved the entire vector $\boldsymbol{z}_M$ rather than a subset, resulting in a bottleneck due to the computational complexity associated with matrix inversion and multiplication.

### 3.2 Orthogonal feedback matrix

To guarantee frequency independence of the energy decay, we adopted the parameterization in (5), where the feedback matrix is computed from an orthogonal matrix $\boldsymbol{U}$. The class of orthogonal matrices meets the unitary condition for losslessness and can be obtained through the parameterization of skew-symmetric matrices described in [22].

At initialization, a random matrix $\boldsymbol{W}$ is constructed. At each optimization step, $\boldsymbol{W}$ is mapped to a skew-symmetric matrix, and the matrix exponential is computed,

$$\boldsymbol{U} = e^{\boldsymbol{W}_{\mathrm{Tr}} - \boldsymbol{W}_{\mathrm{Tr}}^{\top}}\,, \tag{12}$$

where $\boldsymbol{W}_{\mathrm{Tr}}$ is the upper triangular part of $\boldsymbol{W}$ and the operator $e^{(\cdot)}$ denotes the matrix exponential. The mapping in (12) implicitly ensures orthogonality of $\boldsymbol{U}$ and can be used in regular gradient descent optimizers without creating spurious minima [22]. During training, the backward pass differentiates through the parameterization in (12) and $\boldsymbol{W}$ is updated accordingly.

### 3.3 Gain-per-sample

For stable training, the feedback matrix $\boldsymbol{A}$ must have eigenvalues with a modulus less than one, indicating losses. In the lossless case, i.e., $|\lambda_i| = 1$, evaluating $H(\boldsymbol{z}_M)$ becomes infeasible, as the denominator in (3) becomes null.

In our work, we adopted the homogeneous FDN where $\boldsymbol{A}$ is parameterized according to (5), and $\gamma$ is set at initialization to a value lower than one which is constant during optimization. The value of $\gamma$ used during optimization is chosen by examining the connection between the mean damping factor $\overline{\delta}$, used in room acoustics, and the mean spacing of resonance frequencies $\overline{\Delta f}$. To guarantee that the modes are well separated, the mean spacing of resonance frequencies should be larger than the average resonance half-width [23]

$$\overline{\Delta f} \gg \frac{\overline{\delta}}{\pi}\,. \tag{13}$$

In room acoustics, the limiting frequency below which the modes are well-separated is called Schroeder frequency, indicated here as $f_{\mathrm{Schroeder}}$ [24]. This frequency marks the threshold above which an average of at least three modes falls within one resonance half-width. Using the fact that in FDNs the modal frequencies are nearly equally distributed [11], we can derive the limiting average resonance half-width

$$\overline{\Delta f}_{|f=f_{\mathrm{Schroeder}}} = 3\frac{f_s}{\mathcal{M}}\,. \tag{14}$$

We can use the above conditions to determine the minimum value for $T_{60}$ to be used during training

$$T_{60} \gg \frac{\mathcal{M}\ln(10)}{\pi f_s}\,. \tag{15}$$

Increasing the value of $T_{60}$ leads to modes with lower half-widths and greater separation between them. For a target $T_{60}$, the value of $\gamma$ can be derived from (6). However, as $\gamma$ approaches 1, the resonance peaks in the magnitude response become narrow, making it impossible to obtain a flat magnitude response by combining the resonances.

Figure 4 shows the effect of increasing $\gamma$ on the resonance width in a short section of the FDN magnitude response. To enhance the interpretability of the figure, its legend reports the $T_{60}$ values associated with $\gamma$. The sharp peaks visible when $T_{60} = \infty$ ($\gamma = 1$) are significantly smoothed in the other two curves ($\gamma < 1$). For $T_{60} = 0.14$s ($\gamma = 0.999$), the resonance half-widths become excessively large, making it difficult to identify individual resonances.

While (15) provides a lower bound for $\gamma$, defining an upper bound is more complex. Therefore, we recommend treating it as a hyperparameter to be tuned for optimal performance. We ran optimization at different values of $\gamma$, and we observed good convergence when $T_{60} \leq 10$ s. At inference, $\gamma$ serves as a free parameter, enabling the generation of reverberation with arbitrary $T_{60}$ values.

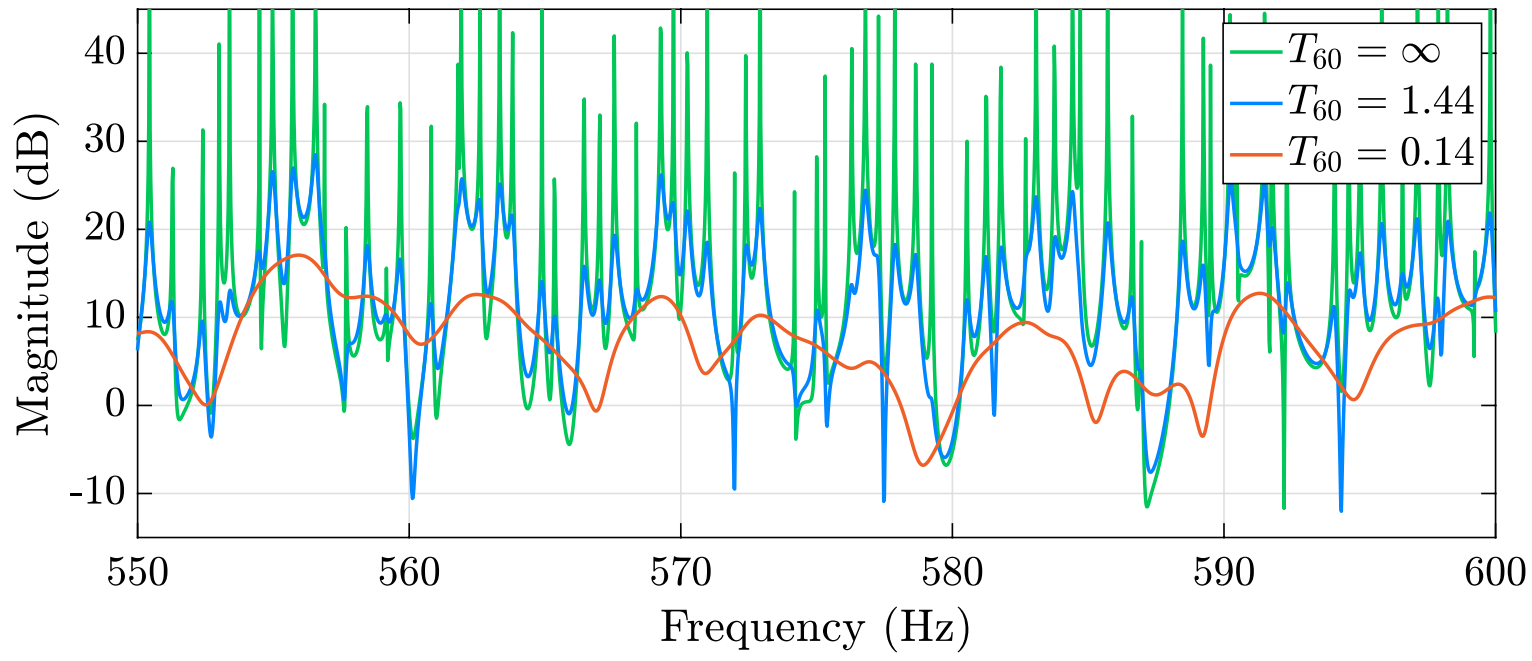

**Fig. 4** Magnitude response of a homogeneous FDN of size $N = 8$ and order $\mathcal{M} = 8944$ with random orthogonal feedback matrix and unitary input/output gains at different values of $T_{60}$ in seconds. In the lossless case, $T_{60} = \infty$, the resonances tend to infinity

### 3.4 Parameter initialization

In stochastic gradient descent, as in many other optimization algorithms, the choice of initial parameter values can affect the algorithm's ability to find the global or a good local minimum of the loss function. We initialize the values of $\boldsymbol{b}$ and $\boldsymbol{c}$ by drawing them from the normal distribution $\mathcal{N}(0, N^{-1})$, and the entries of $\boldsymbol{W}$ from the uniform distribution $\mathcal{U}(-1/\sqrt{N}, 1/\sqrt{N})$. This initialization strategy has empirically yielded good results in our experiments.

The design of the delays is a rather non-trivial task that requires further constraints. To maximize the echo density, the delay lengths should be co-prime [4]. However, the concentration of delays around a certain value may lead to a perceivable strong energy fluctuation over time. Moreover, low-order dependencies, which are integer linear combinations of delays that coincide with other integer linear combinations of delays with small coefficients, can also contribute negatively to the smoothness of the response [9]. To avoid degenerative patterns and ensure a smooth-sounding reverb, we choose delays that are logarithmically distributed co-prime numbers leading to $\mathcal{M} \geq 6000$. The design of delays depends on the specific application, and the gain parameters must be optimized whenever a new set of delay lines is used since they contribute to the system's modal excitation distribution.

### 3.5 Loss function

The network is trained on two losses, $\mathcal{L}_{\text{spectral}}$ and $\mathcal{L}_{\text{sparsity}}$. The spectral loss $\mathcal{L}_{\text{spectral}}$ aims to minimize the frequency-domain mean-squared error between the absolute value of the predicted magnitude response for each channel and the target flat magnitude response. The sparsity loss $\mathcal{L}_{\text{sparsity}}$ penalizes sparseness in the parameter values to encourage density in the time domain. The total loss function is

$$\mathcal{L} = \mathcal{L}_{\text{spectral}}(\boldsymbol{H}(\boldsymbol{z}_\mu)) + \alpha \mathcal{L}_{\text{sparsity}}(\boldsymbol{U}), \tag{16}$$

where $\alpha$ is a weighting term that controls the influence of $\mathcal{L}_{\text{sparsity}}$ on the overall loss. In this work, we set $\alpha$ to 1.

The spectral loss $\mathcal{L}_{\text{spectral}}$ is composed of a channel-wise term (CW) and the summed contribution of all channels (CS)

$$\mathcal{L}_{\text{spectral}} = \mathcal{L}_{\text{spectral}}^{\text{CW}}(\boldsymbol{H}(\boldsymbol{z}_\mu)) + \mathcal{L}_{\text{spectral}}^{\text{CS}}(H(\boldsymbol{z}_\mu)). \tag{17}$$

The individual loss terms are

$$\begin{aligned}
\mathcal{L}_{\text{spectral}}^{\text{CW}}(\boldsymbol{H}(\boldsymbol{z})) &= \frac{1}{\mu} \sum_{z \in z_\mu} \frac{1}{N} \sum_{i=1}^{N} \left(|H_i(z)| - 1\right)^2, \\
\mathcal{L}_{\text{spectral}}^{\text{CS}}(H(z)) &= \frac{1}{\mu} \sum_{z \in z_\mu} \left(|H(z)| - 1\right)^2, \\
\mathcal{L}_{\text{sparsity}}(\boldsymbol{U}) &= \frac{\sum_{i,j} \left|U_{ij}\right| - N\sqrt{N}}{N(1 - \sqrt{N})},
\end{aligned} \tag{18}$$

where $H_i(z)$ is the output of the network's $i^{\text{th}}$ channel computed from the output of the $i^{\text{th}}$ delay line and scaled by $c_i$. Including $\mathcal{L}_{\text{spectral}}^{\text{CW}}$ in (17) is optional, as our tests showed comparable results on the objective metrics with and without it.

The sparsity loss, which is the last term above, is calculated from the sum of the absolute values of the matrix entries, i.e. $\sum_{i,j} \left|U_{ij}\right|$. For orthogonal matrices, its value is bounded between $N$ and $N\sqrt{N}$. To ensure that the loss decreases with increased density and has a dynamic range comparable to $\mathcal{L}_{\text{spectral}}$, we shifted $\sum_{i,j} \left|U_{ij}\right|$ down by $N\sqrt{N}$ and normalized it by $N(1 - \sqrt{N})$. This normalization bounds $\mathcal{L}_{\text{sparsity}}$ between 0 and 1. This sparsity loss is motivated by the fact that echo density increases when the delay lines are highly interconnected. This principle underlies the FDN's structure, distinguishing it from that of parallel comb-filters [3, 20]. Moreover, in our previous work [12] we found that the absence of a term controlling the temporal density of the system's IR may lead the matrix $\boldsymbol{U}$ to converge towards either a diagonal matrix or its permutation. In our previous work [12], the sparseness was controlled by a temporal loss consisting of the ratio of the $\ell_1$ norm to the $\ell_2$ norm of the estimated IR. This required inverse DFT, slowing the optimization steps.

To clarify the impact of the loss function (16) on echo density, Fig. 5 illustrates how the echo density profile, as defined in [25], is influenced by the proposed training, particularly the inclusion of $\mathcal{L}_{\text{sparsity}}$. The left panel in Fig. 5 displays the evolution of the loss and its components over each epoch. The plot shows the losses when an FDN instance with eight delay lines is trained using (16) with $\alpha = 1$, as well as using only $\mathcal{L}_{\text{spectral}}$, i.e. $\alpha = 0$ in (16), indicated in the plot with an asterisk.

In Fig. 5 (left), the terms decay at similar rates, and after six epochs, they begin to converge. The right panel presents the echo density of the FDN at initialization, and after optimization with $\alpha = 1$ and $\alpha = 0$. Both optimizations demonstrate an improvement in echo density, as the curve reaches 1 faster. However, when $\mathcal{L}_{\text{sparsity}}$ is included (i.e., $\alpha = 1$), the echo buildup occurs more rapidly from the early reflections. Moreover, without sparsity control, the magnitude response may converge towards that of a comb filter, leading to a corresponding decrease in IR density, as demonstrated in our previous work [12]. Figure 5 (right), shows also that the proposed $\mathcal{L}_{\text{sparsity}}$ performs comparably to the more computationally expensive sparsity loss term introduced in [12], which is a function of the norm of the impulse response. All evaluated FDNs are initialized with the same parameter values.

To implement the differentiable model and the training iterations, we used the PyTorch library [26]. PyTorch handles parameter adjustments through backpropagation, which involves computing the gradient of the loss function with respect to the given parameters. PyTorch's built-in differentiation engine automatically computes the gradients for any computational graph. Additionally, PyTorch supports complex-valued tensors and backpropagation for real-valued functions of complex tensors, using Wirtinger calculus, which is particularly beneficial for our work in the frequency domain [27].

## 4 Applications

This section showcases how the proposed approach can be effectively employed in practical applications. Specifically, the DiffFDN can be utilized for synthesizing reverberation with a frequency-dependent decay rate and creating dense reverb through the use of attenuation filters and an optimizable scattering feedback matrix.

### 4.1 Frequency-dependent decay control

To generate a natural-sounding room IR that closely matches a measured reference IR, attenuation filters are inserted into a lossless FDN to control the reverberation time as a function of frequency $T_{60}(\omega)$ [18, 28–30]. Using a differentiable FDN optimized for colorlessness as a lossless prototype can address coloration that cannot be effectively mitigated by the attenuation filter, facilitating a reduction in size as well.

The attenuation filter is designed to approximate a target frequency-dependent reverberation time $T_{60}(\omega)$ by achieving a target gain-per-sample $\gamma(\omega)$ in dB as per (6). Attenuation filters are commonly introduced after the delay lines [28]. To simplify the transfer function, at the price of a negligibly lower $T_{60}(\omega)$ accuracy, we place

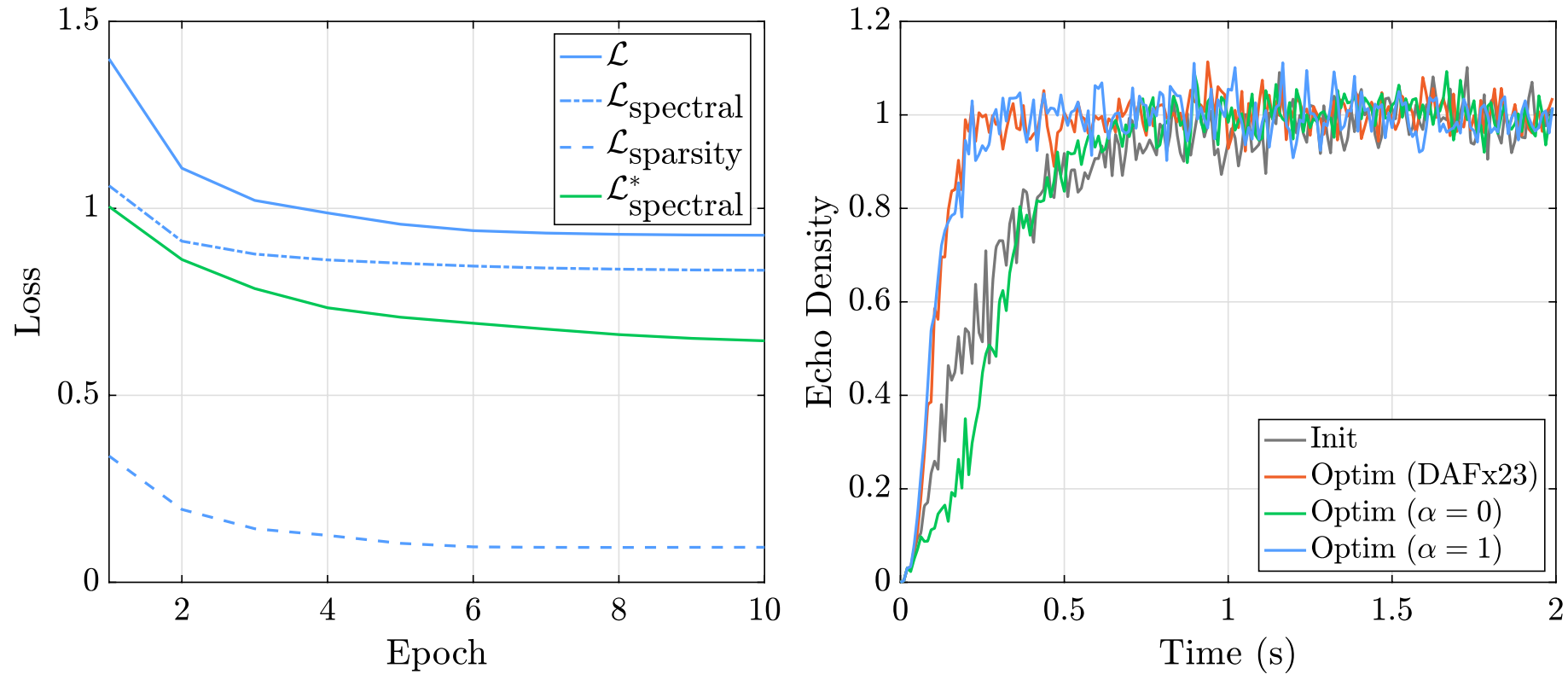

**Fig. 5** Evolution of losses (left) during training of a size 8 FDN with an orthogonal feedback matrix trained with $\alpha = 1$(blue) and $\alpha = 0$ (green, indicated in the legend with the asterisk). On the right, the echo density profile [25] at initialization and after optimization for both cases. For comparison, we included the echo density profile of the FDN trained with the framework in [12] (DAFx23)

the filters in the feedback loop, resulting in the following transfer function:

$$H(z) = \boldsymbol{c}^\top \left[ \boldsymbol{D}_{\boldsymbol{m}}(z)^{-1} - \boldsymbol{U}\boldsymbol{\Gamma}(z) \right]^{-1} \boldsymbol{b} + d\,, \tag{19}$$

where $\boldsymbol{\Gamma}(z)$ is the diagonal attenuation matrix, whose diagonal entries are the delay line specific per-sample attenuation filters $\Gamma_i(z)$. Ideally, $\boldsymbol{\Gamma}(z)$ would be zero-phase, such that the decay rate of the FDN is determined by $|\boldsymbol{\Gamma}(z)|^{1/\boldsymbol{m}}$ [3]. In practical designs, the attenuation filter is not zero-phase, introducing a group delay that is, however, small compared to the delays $\boldsymbol{m}$, hence negligible [3, 18].

We construct the attenuation filters as a graphic equalizer (GEQ) composed of a cascade of eight peaking filters and of a combination of second-order low- and high-shelving filters to control gains at the DC and Nyquist limit, respectively [31]. The peaking filters have center frequencies at one-octave intervals ranging from 63 to 8000 Hz. The low-shelving filter's crossover frequency is set at 46 Hz, and the high-shelving filter's crossover frequency is set at 11,360 Hz. While this filter design shows good performance for reverb synthesis, more accurate designs have been proposed [30, 32], together with their implementation within the DiffFDN [33]. Since this paper focuses on the effect of the DiffFDN itself rather than the choice of filter, we opted to use the simpler eight-band GEQ.

We used DecayFitNet, a neural-network-based approach [34], to estimate each band's $T_{60}$ and decay amplitude $A$ from a reference room IR. DecayFitNet estimates the parameters from the energy decay function modeled as a sum of at most three exponential decays and one noise term. In this work, we assume single slope decay, and we neglect the noise term. Each delay line's output has a per-sample attenuation filter in the feedback path. This filter is designed to compensate for the delay $m_i$ introduced by the delay line by adjusting its target magnitude response

$$\left|\Gamma_i(e^{\jmath\omega})\right| = 10^{m_i \gamma_{\mathrm{dB}}(\omega)/20}\,. \tag{20}$$

The attenuation filters modify the energy of each mode. To align it with the target frequency response envelope, we include a tone correction filter designed as a shared GEQ that matches the initial amplitude provided by the DecayFitNet and placed after the output gains.

### 4.2 Optimizing scattering feedback matrix

A central challenge in the design of FDNs is the generation of sufficient echo density in the IR while maintaining computational efficiency [35]. To make the echo density grow faster in time and reproduce a scattering-like effect, the feedback matrix can be generalized to a filter feedback matrix (FFM) [36] where each entry is an FIR filter. The transfer function of the FDN becomes

$$H(z) = \boldsymbol{c}^\top \left[ \boldsymbol{D}_{\boldsymbol{m}}(z)^{-1} - \boldsymbol{A}(z) \right]^{-1} \boldsymbol{b} + d\,, \tag{21}$$

where $\boldsymbol{A}(z) = \boldsymbol{U}(z)\boldsymbol{\Gamma}(z)$, and $\boldsymbol{U}(z)$ is a FFM, which can be expressed in terms of scalar coefficient matrices $\boldsymbol{U}_0, \boldsymbol{U}_1, \ldots, \boldsymbol{U}_S$, i.e.,

$$\boldsymbol{U}(z) = \boldsymbol{U}_0 + z^{-1}\boldsymbol{U}_1 + z^{-2}\boldsymbol{U}_2 + \ldots z^{-S}\boldsymbol{U}_S\,, \tag{22}$$

where $S$ is the maximum filter order of $\boldsymbol{U}(z)$ if $\boldsymbol{U}_S \neq 0$. The losslessness condition is satisfied if $\boldsymbol{U}(z)$ is paraunitary, i.e., $\boldsymbol{U}(z^{-1})^H \boldsymbol{U}(z) = \boldsymbol{I}$, where $\boldsymbol{I}$ is the identity matrix and $(\cdot)^H$ denotes the complex conjugate transpose [19]. In this work, $\boldsymbol{U}(z)$ is realized as a paraunitary FIR filter using the following factorization:

$$\boldsymbol{U}(z) = \boldsymbol{D}_{\boldsymbol{m}_K}(z)\boldsymbol{U}_K \cdots \boldsymbol{U}_2 \boldsymbol{D}_{\boldsymbol{m}_1}(z)\boldsymbol{U}_1 \boldsymbol{D}_{\boldsymbol{m}_0}(z)\,, \tag{23}$$

where $\boldsymbol{U}_1, \ldots, \boldsymbol{U}_K$ are $N \times N$ unitary matrices and $\boldsymbol{m}_0, \ldots, \boldsymbol{m}_K$ are vectors of $N$ integer delays [36]. In this arrangement, the FFM primarily incorporates $K$ delays and mixing stages into the main FDN loop.

Similar to the scalar feedback matrix, we optimize the unitary matrices and input and output gains to reduce the coloration. We construct the unitary matrices $\boldsymbol{U}_k$ as the coefficients of a $K \times N \times N$ tensor $\boldsymbol{W}_k$ parametrized according to (12) to ensure orthogonality. The delays $\boldsymbol{m}_k$ are fixed, and their values are selected to be relatively small in comparison to the primary delays $\boldsymbol{m}$ to add short-term density. In our design, we used uniformly sampled integer delay values ranging from 1 to a half of the smallest of the primary delays.

### 4.3 Optimizing Householder matrix

An Householder matrix $\boldsymbol{U} = \boldsymbol{I} - 2\boldsymbol{v}\boldsymbol{v}^\top$, where $\boldsymbol{v}$ is a unit vector, gives a simple, yet restricted, parametrization of orthogonal matrices. The Sherman-Morrison formula, i.e.,

$$\left(\boldsymbol{V} + 2\boldsymbol{v}\boldsymbol{v}^\top\right)^{-1} = \boldsymbol{V}^{-1} - \frac{2\boldsymbol{V}^{-1}\boldsymbol{v}\boldsymbol{v}^\top\boldsymbol{V}^{-1}}{1 + 2\boldsymbol{v}^\top\boldsymbol{V}^{-1}\boldsymbol{v}} \tag{24}$$

for an invertible square matrix $\boldsymbol{V}$, helps to avoid the matrix inversion when the Householder parametrization is used. Thus, the FDN transfer function is then written as

$$H(z) = \boldsymbol{c}^\top \left(\boldsymbol{V}(z) + 2\boldsymbol{v}\boldsymbol{v}^\top\right)^{-1} \boldsymbol{b}\,, \tag{25}$$

where $\boldsymbol{V}(z) = \boldsymbol{D}_{\boldsymbol{m}}(z)^{-1} - \boldsymbol{I}$. With this parameterization, the computational complexity required to calculate the FDN transfer function at one frequency point is reduced from $\mathcal{O}(N^3)$ to $\mathcal{O}(N)$.

To ensure homogeneous absorption, $\boldsymbol{\Gamma}$ is included in (25) as

$$H(z) = \boldsymbol{c}^\top \left( \boldsymbol{V}_\gamma(z) + 2\boldsymbol{v}\boldsymbol{v}_\gamma^\top \right)^{-1} \boldsymbol{b}, \quad (26)$$

where $\boldsymbol{V}_\gamma(z) = \boldsymbol{D}_{\boldsymbol{m}}(z)^{-1} - \boldsymbol{\Gamma}$ and $\boldsymbol{v}_\gamma{}^\top = \boldsymbol{v}^\top \Gamma$. With this parametrization, the optimizable parameters become $\boldsymbol{v}$, $\boldsymbol{b}$, and $\boldsymbol{c}$. The input to the sparsity loss remains $\boldsymbol{U}$, which is now defined using the Householder transformation.

## 5 Objective evaluation

The following section presents the FDN configuration and the objective evaluation of the proposed method. The assessment examines the modal excitation distribution.

### 5.1 Evaluation setup

We concentrate on smaller FDNs with limited delay lines, as these are more easily impacted by coloration. Specifically, we consider FDN sizes of $N = 4, 6, 8$ . The values of the delay-line lengths are presented in Table 1. The delay lengths are prime numbers distributed logarithmically. In all configurations, the total number of modes is $8700 < \mathcal{M} < 9000$, which ensures a lack of artifacts related to low modal density [16].

During optimization, we used a sampling rate of $f_s = 48$ kHz and $M = 480000$ frequency points evenly distributed over $[0, \pi]$. The batch size was $\mu = 2000$, requiring 240 steps to complete an epoch. For training, we used 80% of the frequency points, while the remaining 20% were used for validation. We employed Adam optimizer [37] with a learning rate of $\eta = 10^{-3}$. Regarding the gain-per-sample, we found that $\gamma = 0.9999$ leads to stable training and fast convergence. This choice implies $T_{60} = 1.44$s and satisfies (15).

We analyze four configurations for each size, including the differentiable FDN optimized according to our prior work (DAFx23) [12], which uses a random orthogonal feedback matrix. The proposed DiffFDN also incorporates an optimizable orthogonal feedback matrix, with three variations: random orthogonal (DiffFDN-O), Householder (DiffFDN-HH), and scattering (DiffFDN-SCAT).

### 5.2 Computational complexity

The presented optimization utilizes a sparse set of frequency sampling points at each step. This choice aims to decrease the total number of operations, providing a notable reduction compared to the baseline method [12]. During the forward pass, the computational complexity needed to calculate the FDN transfer function at a single frequency point in (11) is $\mathcal{O}(N^3)$. For each epoch, the proposed method (DiffFDN-O) computes the transfer function at $M$ points, where $M$ corresponds to the dataset size. In contrast, the baseline approach (DAFx23) requires calculating a similar number of frequency points at each training step, leading to a linear increase in the number of operations as the dataset size grows, which in [12] is represented by different values of $M$ (e.g., 256 values sampled over [384,000, 480,000]). This is because the number of epochs required for convergence in the baseline is close to that of the presented method. Additionally, in the forward pass of the baseline, the DFT is necessary due to the temporal loss, introducing additional operations. By employing the Householder matrix, we can significantly decrease the computational complexity of (11) to $\mathcal{O}(N)$. Introducing the DFT would raise this complexity to $\mathcal{O}(N \log_2(N))$.

The FDNs also have different computational complexity during operation. For an FDN with $N$ delay lines, each equipped with an octave GEQ, the number of multiply-and-add operations per sample is as follows: $2N$ operations for input and output gains $\boldsymbol{b}$ and $\boldsymbol{c}$, $2N$ operations for the delay lines, $44N$ operations for the attenuation filters, $N^2$ operations for the standard matrix multiplication, and $2N$ for the Householder matrix multiplication. The scattering matrix multiplication with $K$ stages are then $K(N^2 + 2N)$ operations. The comparison of computational complexity for FDN with random orthogonal matrix (RO), DiffFDN-O, DiffFDN-SCAT, and DiffFDN-HH is shown in Table 2 across different matrix sizes.

**Table 1** Delay-line lengths (in samples) for each size *N* of the analyzed FDNs. The lengths are logarithmically distributed prime numbers

| *N* | Delay-line lengths |
|---|---|
| 4 | [1499, 1889, 2381, 2999] |
| 6 | [997, 1153, 1327, 1559, 1801, 2099] |
| 8 | [809, 877, 937, 1049, 1151, 1249, 1373, 1499] |

**Table 2** Number of operations for FDNs of different sizes and matrix types, each incorporating an octave GEQ. In comparison, a typical standard FDN with size $N = 32$ requires 2560 operations

| FDN type | 4 | 6 | 8 |
|---|---|---|---|
| RO | 208 | 324 | 448 |
| DiffFDN-O | 208 | 324 | 448 |
| DiffFDN-HH | 200 | 300 | 400 |
| DiffFDN-SCAT | 288 | 480 | 704 |

**Table 3** Standard deviation of the modal excitation at initialization (Init), after optimization (Optim), and relative decrease from initialization of the tested conditions, at different FDN sizes. Numbers in bold letters indicate the configuration with a greater decrease in standard deviation for each FDN size. Conditions DAFx23 and DiffFDN-O share the same initial parameters. Delay-line lengths are reported in Table 1

| *N* | 4 | | | 6 | | | 8 | | |
|---|---|---|---|---|---|---|---|---|---|
| **FDN type** | **Init** | **Optim** | ↓ | **Init** | **Optim** | ↓ | **Init** | **Optim** | ↓ |
| DAFx23 | 7.7 | 4.7 | −39% | 8.1 | 4.7 | −42% | 7.9 | 4.8 | −39% |
| DiffFDN-O | 7.7 | **4.6** | −40% | 8.1 | **3.4** | −58% | 7.9 | **2.8** | −65% |
| DiffFDN-HH | 9.4 | 6.6 | −30% | 10.2 | 7.7 | −25% | 10.5 | 8.3 | −21% |
| DiffFDN-SCAT | 7.1 | 6.6 | −7% | 7.4 | 7.1 | −4% | 7.6 | 7.3 | −4% |

### 5.3 Modal excitation distribution

To evaluate the improvements in coloration, we analyze the modal excitation distribution before and after the optimization. To obtain the residues $\rho_i$, we compute modal decomposition in (4) using the Ehrlich-Aberth iteration method [11, 38]. The method enables solving the eigenvalue problem for large delays $\boldsymbol{m}$ by finding an approximation to the eigenvalues based on the polynomial matrix formulation of the FDN [38].

Table 3 shows the standard deviation of the modal excitation distribution in dB, $20\log_{10}|\rho_i|$, for each tested configuration. The numbers refer to the mean of the standard deviation over 100 iterations, each with a unique set of randomly sampled initial parameters. Each optimization run is 20 epochs long for DAFx23 and DiffFDN-O, and 30 epochs long for DiffFDN-HH and DiffFDN-SCAT. To run the objective test, we used FLAMO, a PyTorch library for the optimization of linear time-invariant systems in frequency domain [39]. The code is available in a dedicated branch of the project repository[1]. In each scenario, the standard deviation decreases, signifying a reduction in coloration. However, based on this test alone, it is challenging to predict the overall trend across different FDN sizes.

To assess the impact of the changes to our previous optimization framework [12]-particularly the addition of the sparsity loss function and sparse frequency sampling-we included the original optimization framework (DAFx23) for comparison. Each set of parameters used at initialization is shared between DAFx23 and DiffFDN-O conditions.

For $N = 4$, the proposed method shows results comparable to our previous work (DAFx23). Even though it is likely impossible to perceive a difference of 0.1 dB in standard deviation, the more significant improvement lies in the reduction of operation count. For larger FDNs, the proposed method (DiffFDN-O) shows a narrower distribution at convergence.

Due to a different parametrization of the feedback matrix, in DiffFDN-HH and DiffFDN-SCAT the distributions of the initial parameters differ from DAFx23 and DiffFDN-O. After optimization, the standard deviation reduces in both cases. Among all tested configurations, the DiffFDN-HH case exhibited the highest standard deviation, likely due to its more restrictive parametrization. The DiffFDN-SCAT case also showed a lower degree of improvement. This could be attributed to the higher complexity of the problem, suggesting potential enhancement through fine-tuning the optimization framework. Moreover, our experiments showed that with more complex structures, like the DiffFDN-SCAT, the optimization performance is more influenced by the initial parameter values, potentially causing convergence to poorly sounding local minima.

The histograms in Fig. 6 illustrate the distribution of modal excitation at the start (Init) and end of the optimization process (Optim) for a single instance of DiffFDN-O, across all analyzed sizes. The optimization reduces the loudest modes, leading to a shift toward a narrower overall excitation distribution. Figure 7 shows the effect on their magnitude response. Similarly to Fig. 2, only a narrow section of the response is displayed. While resonances are inherent to the FDN structure, in the optimized responses, their prominence has been reduced, and their peak gain shows fewer fluctuations across frequencies. The effect of these changes on perceived coloration is further evaluated through a subjective test.

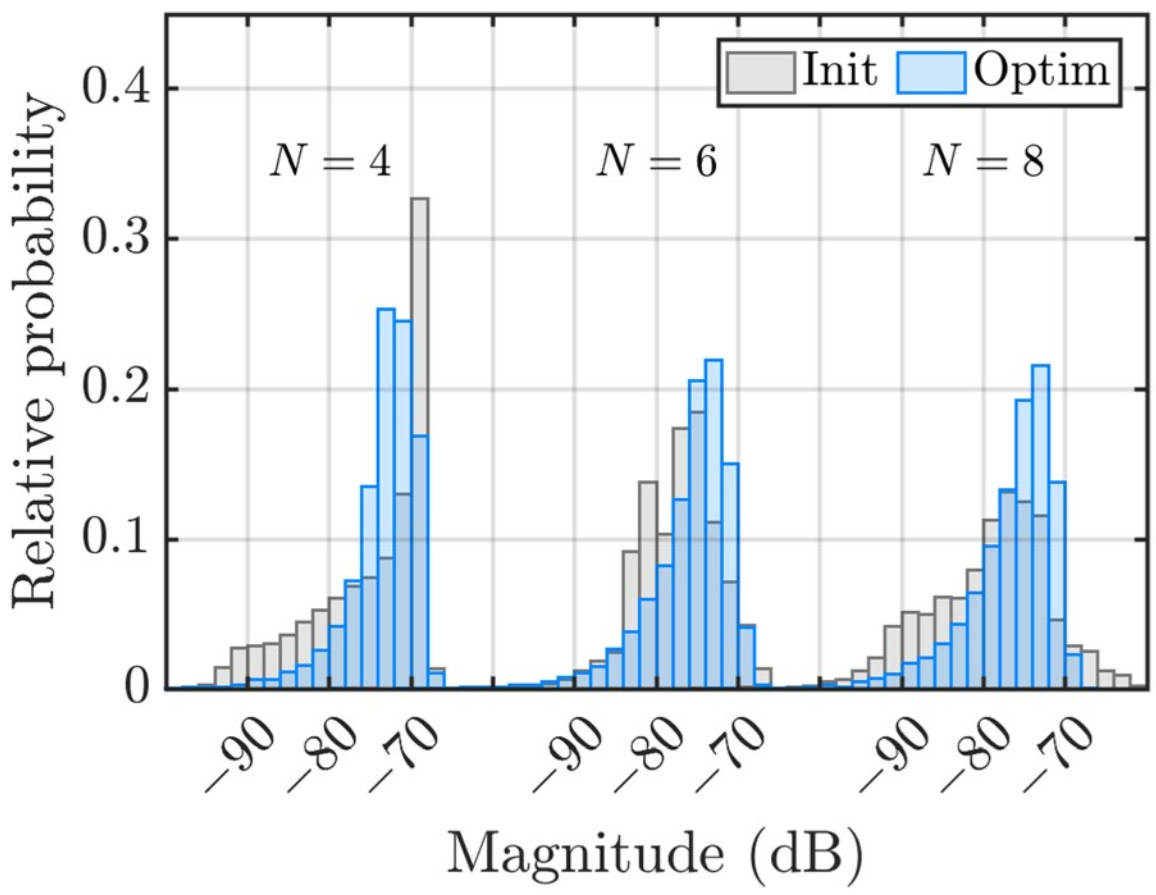

**Fig. 6** Distribution of modal excitations of FDNs with sizes $N = 4, 6,$ and $8$ before (Init) and after the optimization (Optim). The optimization led to a decrease in the loudest modal excitations and to a narrower distribution in each case, as desired

[1] https://github.com/gdalsanto/diff-fdn-colorless/

## 6 Perceptual evaluation

To further validate our results, a formal listening test was conducted to evaluate the perceived coloration in the IRs of the optimized FDN and quantify the improvement from IRs produced with the initial parameters.

### 6.1 Listening test procedure

The test followed the Multiple Stimuli with Hidden Reference and Anchor (MUSHRA) recommendation [40], and it was carried out using the web audio API-based experiment software webMUSHRA developed by International Audio Laboratories Erlangen [41].

On each page, the listening test compared five FDN configurations against a reference. Similarly to the objective evaluation, the test items included three FDN sizes ($N = 4$, 6, and 8). The configurations were evaluated on three different reverberation conditions, each covering a dedicated section of the test. The tested reverberation conditions are *lossless*, *homogeneous decay*, and *frequency-dependent decay*. At the beginning of each part, a training page was presented to familiarize the subjects with the sound samples. Adjustments to the overall loudness were allowed during the initial training page and it was maintained constant throughout the remainder of the test.

During the evaluation, participants rated the similarity between each presented item and the reference sound using a scale ranging from 0 to 100. On each page, six IRs were assessed, including an anchor and the hidden reference. To encourage subjects to assess samples based on coloration rather than subtle temporal features, the hidden reference differed from the actual reference as a distinct instance of the generated signal while maintaining its statistical properties.

In the part testing the *lossless* condition, we evaluated late non-decaying reverb. The reference for this part was a 3-s segment of white Gaussian noise, known to be highly colorless. The IRs of the FDNs were extracted starting from the mixing time, thereby eliminating the initial echo buildup from the evaluation. The section testing *homogeneous decay*, focused on frequency-independently decaying IRs, with a $T_{60}$ value set at 2 s for all frequencies. A corresponding exponential decaying envelope with the same $T_{60}$ value was applied to the reference white Gaussian noise. The part testing *frequency-dependent decay* involved decaying IRs with a fixed frequency-dependent $T_{60}(\omega)$ across conditions. The reference in this case was the IRs of an FDN of size $N = 64$, ensuring a sufficiently smooth response. The frequency-dependent attenuation filter was designed following the steps outlined in Sect. 4.1.

The test evaluated the coloration of DiffFDN-O, DiffFDN-SCAT, and DiffFDN-HH. All tested items of a page share the same delay-line lengths and reverberation condition, and only the feedback matrix and the input and output gains were altered. The FDN implementation of the Schroeder series allpass reverberator (SH) was the anchor, whereas the RO condition acted as a baseline. For the RO, the initial values of DiffFDN-O were used. The direct gain $d$ was set to zero in all cases. As the primary contribution of this work is focused on improving

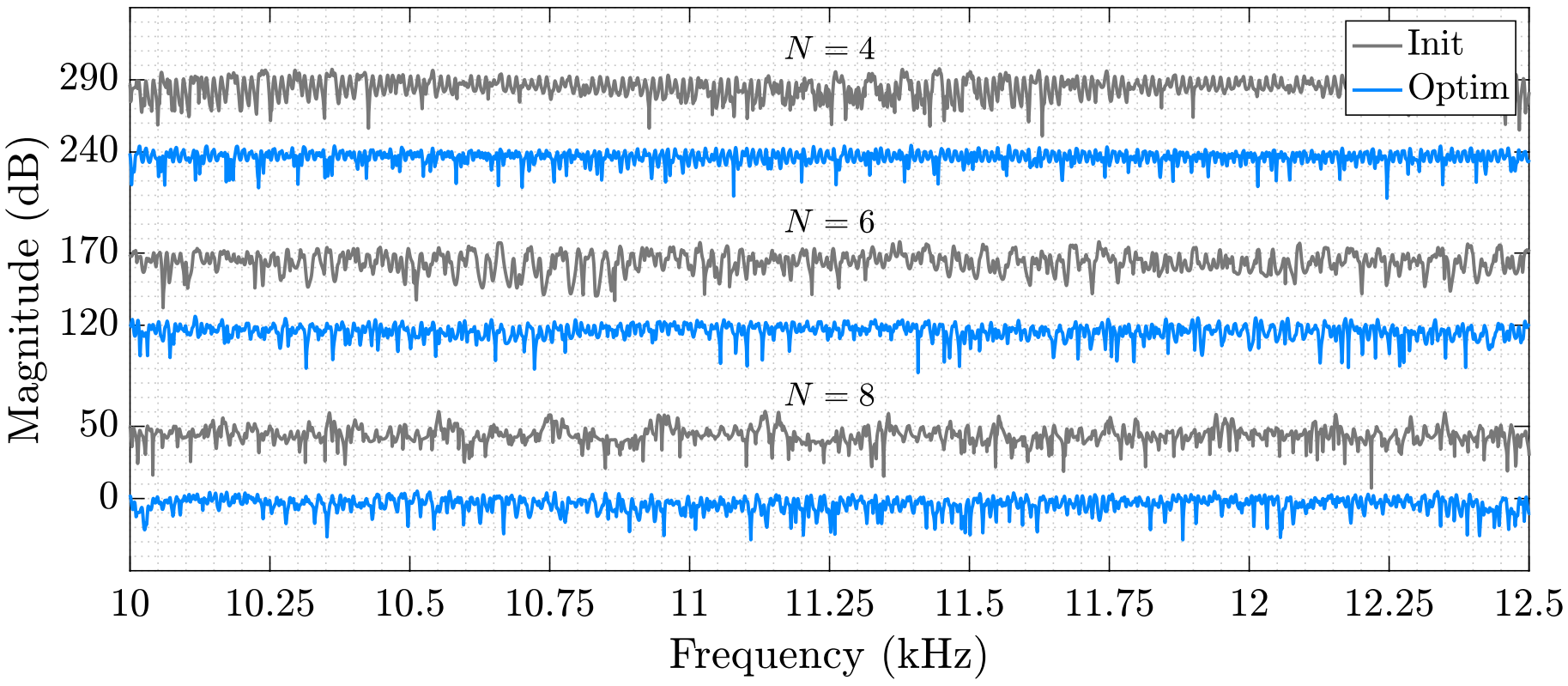

**Fig. 7** Section of the magnitude response of the FDNs in Fig. 6. The magnitude of responses at initialization (Init) and at the end of optimization (Optim) was shifted for enhanced visual clarity. During the optimization, the resonances were reduced and more evenly distributed

training efficiency compared to our previous work, we excluded the DiffFDN trained with the framework presented in [12] Adding this condition to the test would constitute a stimulus perceptually nearly identical to DiffFDN-O, possibly skewing the results. In total, there were 18 listening test pages with 6 stimuli each. The loudness of each IR was normalized to ensure consistency across conditions.

The experiment was conducted in a sound-insulated booth at the Aalto Acoustics Lab, with participants wearing Sennheiser HD 650 headphones. The final items were presented to 14 listeners. The average age of the participants analyzed was 29.5 years, with a standard deviation of 4.72, and none of them reported any hearing impairments. All participants, except one, were either students or employees of the Aalto University Acoustics Lab and had previous experience with the MUSHRA test. The results from three subjects were removed from the following analysis because they were detected to be outliers according to the MUSHRA recommendation [40].

The box plots presented in Fig. 8 illustrate the outcomes of the listening test. In each box, the median is represented by the central mark, while the lower and upper edges indicate the 25th and 75th percentiles, respectively. The whiskers extend to encompass the most extreme data points not identified as outliers, with any outliers plotted separately. The shaded regions surrounding the medians facilitate the comparison of sample medians across various box charts. Non-overlapping shaded regions signify differing medians between the compared box charts at the 5% significance level, assuming a normal distribution.

Despite excluding the reference and anchor conditions, the data, as revealed by the Shapiro-Wilk test [42], deviated from a normal distribution. Additionally, the Wilcoxon signed-rank test [43] was employed to assess the score distributions for each pair of conditions within each page. To address multiple comparisons (15 hypotheses

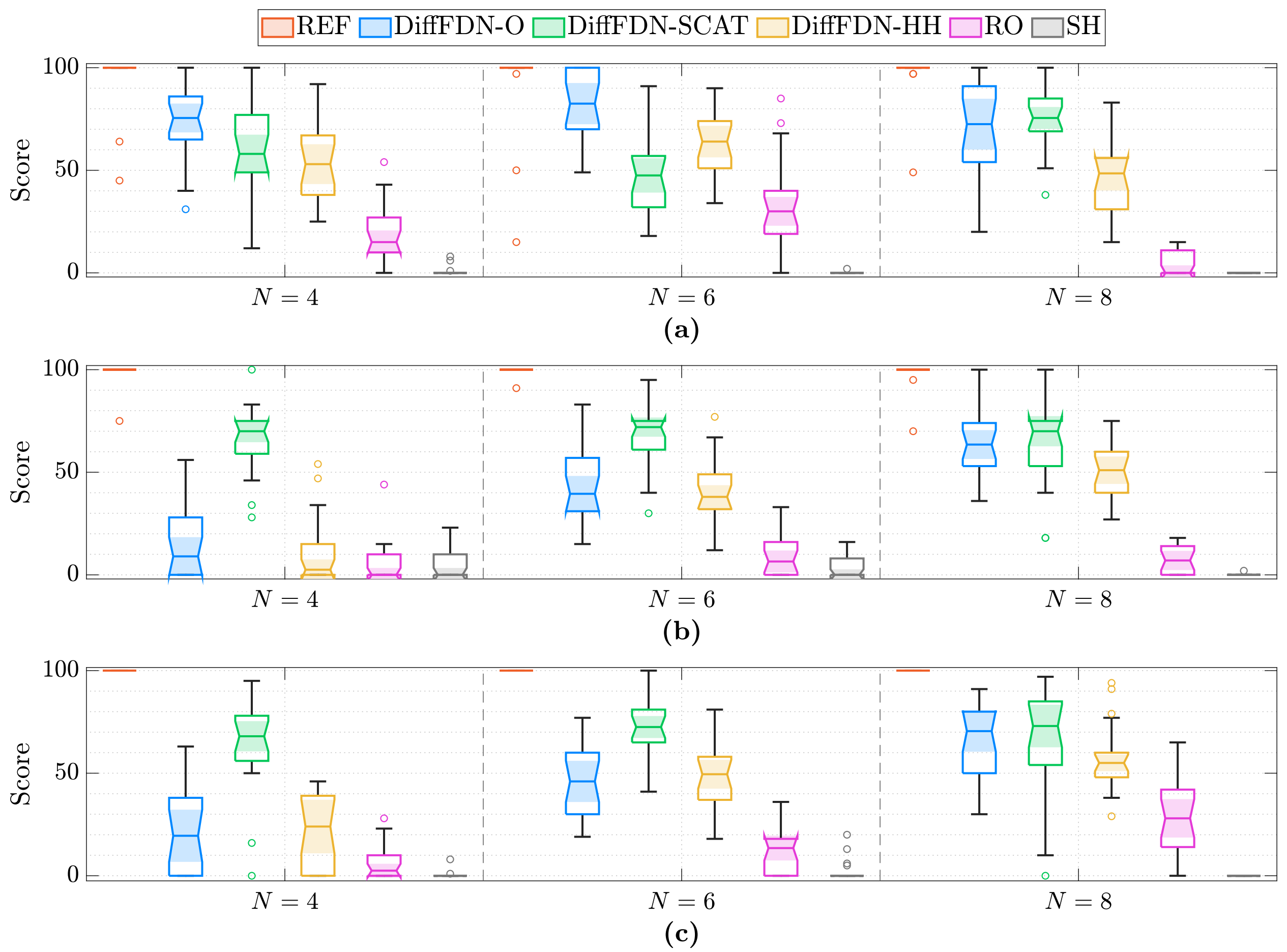

**Fig. 8** Listening test results on the analyzed reverberation conditions: **a** lossless, **b** homogeneous decay, and **c** frequency-dependent decay. The three DiffFDN conditions correspond to FDN configurations trained using the proposed optimization framework

per page), we applied the Bonferroni method to adjust the alpha level.

The $p$-values for most paired conditions indicate significant differences among all pairs of results. However, there are a few exceptions observed in the distribution of DiffFDN-O and DiffFDN-HH. In these cases, the $p$-values exceed the alpha level across most conditions with decaying modes suggesting a convergence in the scores assigned to the two conditions. Moreover, DiffFDN-O and DiffFDN-HH reject the null hypothesis when paired with DiffFDN-SCAT in the lossy cases for $N = 8$, suggesting that for larger FDN sizes, the performance improvement provided by DiffFDN-SCAT diminishes.

The IRs of the optimized FDNs always show a median greater than the initial condition (RO), proving the effectiveness of the optimization. In the late reverb analysis, Fig. 8a, the DiffFDN-O has median ratings of 75.5, 82.5, and 72.5, respectively, for increasing FDN size. Only for $N = 8$, the FDN with scattering matrix received ratings higher than DiffFDN-O with a median of 75.5.

In the *homogeneous decay* and *frequency-dependent decay* tests the full IRs are taken into account. The scores given to the DiffFDN-SCAT condition suggest that the scattering matrix contributes to the smoothness of the early reflections as well as the reverberation tail. In general, spectral flatness is associated with temporal smoothness. The scattering feedback matrix increases echo density during early reflections, potentially suggesting a less colored response, as shown in Fig. 8b, c. This effect becomes more noticeable for smaller FDNs, and for $N = 8$, the medians of the results begin to converge among optimized FDNs. In [33], the authors demonstrated in a listening test that, by utilizing the scattering feedback matrix, the DiffFDN can outperform neural network-based approaches when synthesizing real room IRs, highlighting the benefit of faster echo density build-up.

The confidence intervals for the results of all tests are relatively large. Although the objective results in Table 3 indicate a limited reduction in modal excitation standard deviation, the optimized Householder matrix was rated relatively close to the DiffFDN-O in the listening test. The distribution of modal excitation is one quantity that can aid in deducing the coloration of the reverberator. However, this is a complicated concept, as evidenced by the results presented in the listening test.

## 7 Conclusion

This work introduces an optimization method for designing artificial reverberation using a tiny differentiable feedback delay network, or DiffFDN. The method focuses on achieving spectral flatness and temporal density by optimizing the feedback matrix and the input and output gains. Emphasis has been placed on improving the computational efficiency of the method without compromising its performance. By incorporating attenuation filters and an optimizable scattering feedback matrix, this method can be further enhanced, presenting an efficient option for synthesizing natural room impulse responses.

Favorable results were obtained in this study with a tiny DiffFDN employing as few as four delay lines. In objective evaluations, we demonstrated that the proposed method reduces the width of the modal excitation distribution, thereby decreasing the prominence of the loudest modes. Listening test results further confirmed the success of this approach in attenuating artifacts commonly associated with undesired coloration, making the DiffFDN close to an ideally colorless, smooth late reverberation.

**Acknowledgements**
Not applicable.

**Authors' contributions**
GDS performed the methodology, analysis, validation, and wrote the original draft. KP performed the conceptualization, writing, and project supervision. SJS performed conceptualization, methodology, analysis, writing, and project supervision. VV performed project supervision, review, and editing.

**Funding**
The Aalto University School of Electrical Engineering funded the work of the first author.

**Data availability**
The PyTorch implementation of the proposed method, together with the code used to generate the figures, can be found on the online repository[1]. Listening test stimuli examples can be found on a dedicated website[2].

## Declarations

**Competing interests**
The authors declare that they have no competing interests.

**References**

1. V. Välimäki, J.D. Parker, L. Savioja, J.O. Smith, J.S. Abel, Fifty years of artificial reverberation. IEEE Trans. Audio Speech Lang. Process. **20**(5), 1421–1448 (2012)
2. M.R. Schroeder, B.F. Logan, "Colorless" artificial reverberation. IRE Trans. Audio **6**, 209–214 (1961)
3. J.M. Jot, A. Chaigne, in *Proceedings of the Audio Engineering Society Convention 90*, Digital delay networks for designing artificial reverberators (Paris, France, 1991)
4. S.J. Schlecht, Feedback delay networks in artificial reverberation and reverberation enhancement (Ph.D. thesis, Friedrich-Alexander-Universitaet Erlangen-Nuernberg, Germany, 2018)
5. C. Kirsch, T. Wendt, S. Van De Par, H. Hu, S.D. Ewert, Computationally-efficient simulation of late reverberation for inhomogeneous boundary conditions and coupled rooms. J. Audio Eng. Soc. **71**(4), 186–201 (2023)

[2] http://research.spa.aalto.fi/publications/papers/eurasip-colorless-fdn/.

6. J. Herre, S. Disch, MPEG-I immersive audio — reference model for the virtual/augmented reality audio standard. J. Audio Eng. Soc. **71**(5), 229–240 (2023)
7. D. Rocchesso, J.O. Smith, Circulant and elliptic feedback delay networks for artificial reverberation. IEEE Trans. Speech Audio Process. **5**(1), 51–63 (1997)
8. S.J. Schlecht, E.A. Habets, On lossless feedback delay networks. IEEE Trans. Signal Proc. **65**(6), 1554–1564 (2016)
9. S.J. Schlecht, E.A. Habets, Feedback delay networks: Echo density and mixing time. IEEE/ACM Trans. Audio Speech Lang. Process. **25**(2), 374–383 (2016)
10. J. Heldmann, S.J. Schlecht, in *Proceedings of the 24th International Conference on Digital Audio Effects (DAFx20in21)*, The role of modal excitation in colorless reverberation (Online, 2021), pp. 206–213
11. S.J. Schlecht, E.A. Habets, Modal decomposition of feedback delay networks. IEEE Trans. Signal Proc. **67**(20), 5340–5351 (2019)
12. G. Dal Santo, K. Prawda, S. Schlecht, V. Välimäki, in *Proceedings of the 26th International Conference on Digital Audio Effects (DAFx23)*, Differentiable feedback delay network for colorless reverberation (Copenhagen, Denmark, 2023), pp. 244–251
13. A.I. Mezza, R. Giampiccolo, E. De Sena, A. Bernardini, Data-driven room acoustic modeling via differentiable feedback delay networks with learnable delay lines. EURASIP J Audio Speech Music Process. **2024**(1), 51 (2024)
14. A.I. Mezza, R. Giampiccolo, A. Bernardini, in *Proceedings of the 27th International Conference on Digital Audio Effects (DAFx24)*, Modeling the frequency-dependent sound energy decay of acoustic environments with differentiable feedback delay networks (Guildford, UK, 2024), pp. 238–245
15. R. Giampiccolo, A.I. Mezza, A. Bernardini, in *Proceedings of the 27th International Conference on Digital Audio Effects (DAFx24)*, Differentiable MIMO feedback delay networks for multichannel room impulse response modeling (Guildford, UK, 2024), pp. 278–285
16. M. Karjalainen, H. Järveläinen, in *Proceedings of the 111th Audio Engineering Society Convention*, More about this reverberation science: Perceptually good late reverberation (New York, NY, USA, 2001)
17. J.M. Jot, in *Proceedings of the Audio Engineering Society Convention 139*, Proportional parametric equalizers—Application to digital reverberation and environmental audio processing (New York, NY, USA, 2015)
18. S.J. Schlecht, E.A. Habets, in *Proceedings of the 20th International Conference on Digital Audio Effects (DAFx17)*, Accurate reverberation time control in feedback delay networks (Edinburgh, 2017), pp. 337–344
19. M.A. Gerzon, Unitary (energy-preserving) multichannel networks with feedback. Electron. Lett. **11**(12), 278–279 (1976)
20. J. Stautner, M. Puckette, Designing multi-channel reverberators. Comput. Music. J. **6**(1), 52–65 (1982)
21. J.A. Moorer, About this reverberation business. Comput. Music. J. **3**(2), 13–28 (1979)
22. M. Lezcano-Casado, D. Martínez-Rubio, in *Proceedings of the International Conference on Machine Learning*, Cheap orthogonal constraints in neural networks: A simple parametrization of the orthogonal and unitary group (Long Beach, CA, USA, 2019), pp. 3794–3803
23. H. Kuttruff, Room Acoustics, 5th edn. (CRC Press, 2009)
24. H. Kuttruff, Eigenschaften und Auswertung von Nachhallkurven. Acta Acustica U. Acustica **8**(4), 273–280 (1958)
25. J.S. Abel, P. Huang, in *Proceedings of the Audio Engineering Society Convention 121*, A simple, robust measure of reverberation echo density (Audio Engineering Society, 2006)
26. A. Paszke, S. Gross, S. Chintala, G. Chanan, E. Yang, Z. DeVito, Z. Lin, A. Desmaison, L. Antiga, A. Lerer, in *31st Neural Information Processing Systems Automatic differentiation Workshop* (Long Beach, CA, USA, 2017)
27. PyTorch complex numbers. https://pytorch.org/docs/stable/complex_numbers.html. Accessed 23 Aug 2024
28. J.M. Jot, in *Proceedings of IEEE International Conference on Acoustics, Speech, and Signal Processing*, vol. 2. An analysis/synthesis approach to real-time artificial reverberation (San Francisco, CA, USA, 1992), pp. 221–224
29. K. Prawda, S.J. Schlecht, V. Välimäki, in *Proceedings of the 22nd International Conference on Digital Audio Effects*, Improved reverberation time control for feedback delay networks (Birmingham, UK, 2019), pp. 1–7
30. V. Välimäki, K. Prawda, S.J. Schlecht, Two-stage attenuation filter for artificial reverberation. IEEE Signal Proc. Lett. **31**, 391–395 (2024)
31. V. Välimäki, J.D. Reiss, All about audio equalization: Solutions and frontiers. Appl. Sci. **6**(5), 25–70 (2016)
32. J. Liski, V. Välimäki, in *Proceedings of the 20th International Conference on Digital Audio Effects*, The quest for the best graphic equalizer (Edinburgh, UK, 2017), pp. 95–102
33. G. Dal Santo, B. Alary, K. Prawda, S. Schlecht, V. Välimäki, in *Proceedings of the 27th International Conference on Digital Audio Effects (DAFx24)*, RIR2FDN: An improved room impulse response analysis and synthesis (University of Surrey, 2024), pp. 230–237
34. G. Götz, R. Falcón Pérez, S.J. Schlecht, V. Pulkki, Neural network for multi-exponential sound energy decay analysis. J. Acoust. Soc. Am. **152**(2), 942–953 (2022)
35. S.J. Schlecht, E.A. Habets, in *Proceedings of the IEEE Workshop on Applications of Signal Processing to Audio and Acoustics (WASPAA)*, Dense reverberation with delay feedback matrices (New Paltz, NY, USA, 2019), pp. 150–154
36. S.J. Schlecht, E.A. Habets, Scattering in feedback delay networks. IEEE/ACM Trans Audio Speech Lang. Process. **28**, 1915–1924 (2020)
37. D.P. Kingma, J. Ba, Adam: A method for stochastic optimization (2014). arXiv:1412.6980
38. D.A. Bini, V. Noferini, Solving polynomial eigenvalue problems by means of the Ehrlich-Aberth method. Linear Algebra Appl. **439**(4), 1130–1149 (2013)
39. G. Dal Santo, G.M. De Bortoli, K. Prawda, S.J. Schlecht, V. Välimäki, FLAMO: An open-source library for frequency-domain differentiable audio processing (2024). arXiv:2409.08723
40. International Telecommunication Union, Method for the subjective assessment of intermediate quality level of audio systems. Recommendation ITU-R BS.1534–3 (2015)
41. M. Schoeffler, S. Bartoschek, F.R. Stöter, M. Roess, S. Westphal, B. Edler, J. Herre, WebMUSHRA—a comprehensive framework for web-based listening tests. J. Open Res. Softw. **6**(1), 1–34 (2018)
42. S.S. Shapiro, M.B. Wilk, An analysis of variance test for normality (complete samples). Biometrika **52**(3/4), 591–611 (1965)
43. F. Wilcoxon, Individual comparisons by ranking methods. Biom. Bull. **1**(6), 80–83 (1945)

## Publisher's Note